\documentstyle[12pt,aasms4,epsf]{article}
\begin{document}

\newcommand{\etal}{ {et al.}}

\newcommand{\ce}[1]{\centerline{\bf{#1}} \vskip .2in}
\newcommand{\cen}[1]{\centerline{#1}}

\newcommand{\be}{\begin{equation}}
\newcommand{\en}{\end{equation}}

\def\loe{\lower 0.6ex\hbox{${}\stackrel{<}{\sim}{}$}}
\def\goe{\lower 0.6ex\hbox{${}\stackrel{>}{\sim}{}$}}

\def\ggg{$\gamma$}

%
\def\jref#1 #2 #3 #4 {{\par\noindent \hangindent=3em \hangafter=1
      \advance \rightskip by 0em #1, {\it#2}, {\bf#3}, #4.\par}}
\def\rref#1{{\par\noindent \hangindent=3em \hangafter=1
      \advance \rightskip by 0em #1.\par}}

\def\p{\phantom{1}}
\def\pmu{\mox{$^{-1}$}}
\def\ApJ{{\it Ap.\,J.\/}}
\def\ApJL{{\it Ap.\,J.\ (Letters)\/}}
\def\ApJS{{\it Astrophys.\,J.\ Supp.\/}}
\def\AJ{{\it Astron.\,J.\/}}
\def\AAL{{\it Astr.\,Astrophys.\ Letters\/}}
\def\AAS{{\it Astr.\,Astrophys.\ Suppl.\,Ser.\/}}
\def\MN{{\it Mon.\,Not.\,R.\,Astr.\,Soc.\/}}
\def\Na{{\it Nature \/}}
\def\SAIt{{\it Mem.\,Soc.\,Astron.\,It.\/}}
\def\BGD{\begin{description}}
\def\EDD{\end{description}}
\def\BGF{\begin{figure}}
\def\EDF{\end{figure}}
\def\BGC{\begin{center}}
\def\EDC{\end{center}}
\def\BGT{\begin{tabular}}
\def\EDT{\end{tabular}}
\def\BGE{\begin{equation}}
\def\EDE{\end{equation}}
\def\REFFF{\par\noindent\hangindent 20pt}
\def\DS{\displaystyle}
\def\kms{km^s$^{-1}$}
\def\sbu{mag^arcsec${{-2}$}}
\def\e{\mbox{e}}
\def\dex{\mbox{dex}}
\def\L{\mbox{${\cal L}$}}
 

\newcommand{\porb}{ P_{orb} } 
\newcommand{\Po}{$ P_{orb} \su$}
\newcommand{\pdot}{$ \dot{P}_{orb} \,$}

\newcommand{\pot}{$ \dot{P}_{orb} / P_{orb} \su $}
\newcommand{\s}{ \\ [.15in] }

\newcommand{\mm}{$ \dot{m}$ }
\newcommand{\mdot}{$ |\dot{m}|_{rad}$ }

\newcommand{\myr}{ \su M_{\odot} \su \rm yr^{-1}}
\newcommand{\msol}{\, M_{\odot}}
\newcommand{\ppp}{ \dot{P}_{-20} }

\newcommand{\ci}[1]{\cite{#1}}
\newcommand{\bb}[1]{\bibitem{#1}}

\newcommand{\ch}[1]{\vskip .3in \noindent {\bf #1} \para}

\newcommand{\cms}{ \rm \, cm^{-2} \, s^{-1} }

\newcommand{\nn}{\noindent}

%
%

\renewcommand{\thebibliography}[1]{
  \list
  {[\arabic{enumi}]}{\settowidth\labelwidth{[#1]}\leftmargin\labelwidth
    \advance\leftmargin\labelsep
    \usecounter{enumi}}
    \def\newblock{\hskip .11em plus .33em minus .07em}
    \parsep -2pt
    \itemsep \parsep
    \sloppy\clubpenalty4000\widowpenalty4000
    \sfcode`\.=1000\relax}

\newcommand{\pdott}{ \left( \frac{ \dot{P}/\dot{P}_o}{P_{1.6}^{3}} \right)}

\newcommand{\asca}{{\it ASCA} }
\newcommand{\psr}{PSR~B1259-63 }
\newcommand{\psrp}{PSR~B1259-63} 

\newcommand{\syn}{synchrotron }

\newcommand{\flux}{\rm \, erg \, cm^{-2} \, s^{-1}}
\newcommand{\ergs}{\rm \, erg \,  s^{-1}}

\def\grbj{GRB~970111 }
\def\grbjp{GRB~970111}
\def\grbf{GRB~970228 }
\def\grbfp{GRB~970228}

\vspace*{.4in}
\nn


 \cen{\Large \bf  X-Ray Afterglows from  Gamma-Ray Bursts}

\vskip .1in
\cen{\large M. Tavani}
\cen{\it Columbia Astrophysics Laboratory, Columbia University, New York, NY 10027}
\cen{\it IFCTR-CNR, via Bassini 15, Milano (Italy)}

\vskip .4in
\ce{Abstract}

We consider possible  interpretations of the
recently detected X-ray afterglow  from the
gamma-ray burst source  \grbfp.
 Cosmological and Galactic models of gamma-ray bursts predict
different flux and spectral evolution of X-ray afterglows.
We show that  models based on adiabatic expansion of relativistic
forward shocks 
require very efficient particle
energization  or post-burst re-acceleration 
during the expansion.
Cooling neutron star models predict a very distinctive 
spectral and flux evolution  that can be tested in current X-ray data.

\vskip 1.in
\cen{Subject Headings: gamma rays: bursts, theory; X-rays: general}

\vskip .4in
\cen{March 14, 1997}
\cen{Submitted to the {\it Astrophysical Journal Letters}: March 18, 1997}

\newpage

\section{Introduction}

The discovery by SAX of an X-ray afterglow from \grbf (Costa \etal 1997b,c,
hereafter C97b,c)
provides crucial information for the understanding of gamma-ray bursts (GRBs).
This detection was possible because of a relatively rapid response in
pointing to a GRB error box within a few hours.
The results of two `fast' Target of Opportunity (TOO) SAX pointings
 of GRB error boxes are currently available.
An observation within 16~hrs  of the $\sim 15 \, \rm arcmin^2$
 error box of \grbj (Hurley \etal\ 1997a) 
 did not  detect X-ray
emission above a flux of $10^{-13} \flux$ (2-10~keV band) 
(Costa \etal\ 1997b).
On the contrary, a pointing within 8 hours at the
$\sim 6 \, \rm arcmin^2$  error box of \grbf (Hurley \etal\ 1997b)
resulted in the remarkable discovery
of transient X-ray emission from a previously unknown source (C97c,
Boller \etal\ 1997).
 The average  X-ray  flux
level (2-10~keV band) of the first TOO  pointing at the \grbf error 
box is $\sim 3\times 10^{-12} \flux$.
A  second observation
carried out $\sim$3 days later
 detected
the same X-ray source at a substantially lower flux  near 
$10^{-13} \flux$ (C97c).
In this {\it Letter} we consider theoretical models that  can
reproduce the observed decay in X-ray flux by  a factor of $\sim 30$
between the first (at $t\sim 8$~hrs, with $t$ the elapsed time after the 
burst trigger) and second (at $t \sim 3.6$~days $\sim 86.6$~hrs)
SAX observations of \grbfp.
It is also interesting to note the absence\footnote{
Sources (a) and (b) detected by SAX in the WFC error box of \grbj
(Butler \etal\ 1997)
are outside the refined error box of in 't Zand \etal\ (1997).}
 of  a relatively strong X-ray
source (at flux level near $10^{-12} \flux$)  $\sim 16$~hrs after the
\grbj trigger. 
The peak intensities of \grbj and \grbf as detected by the GRBM instrument
in the 60-600 keV band
are comparable within a factor of $\sim 2$ (C97a,b).
We infer that
temporal evolutions of X-ray afterglows from GRBs differ among bursts,
and suggest that GRB  models with no universal afterglow evolution are most likely.



\section{Cosmological models}

Coalescence of compact stars, failed supernovae, new-born spinars or
special jet phenomena  in distant galaxies
 can give rise to a rapidly expanding relativistic fireball 
dissipating its energy in internal and external
shocks. To avoid pair opacity effects, these models require
a large value of the relativistic 
bulk Lorentz factor $\Gamma \sim 10^2-10^3$ characterizing 
a fast-moving radiation front. 
  The observed photon energies are upscattered by $\Gamma$
compared to the ones  produced in the comoving frame.
 The observed spectral energy flux $F_{\nu}$ is 
$\Gamma^5$ times its comoving value, a factor of $\Gamma^2$ from the 
increasing solid angle from the apparent
transverse dimension of the fireball ($r_{\perp} = c \, t \, \Gamma$), and 
a factor of $\Gamma^3$ from relativistic beaming (Rees 1966). 
Adiabatic expansion applies
to  fireball emission occurring within a 
time $\sim 10^3-10^4$ times longer than the burst $T_{90}$
duration ($T_{90} \goe 20$~s for \grbj and \grbfp,
C97a,b). The initial burst may be
produced by 
`internal' or `external' shocks  for an initial 
value of $\Gamma = \Gamma_o \sim 10^2-10^3$ (
e.g., M\'esz\'aros \& Rees 1997, hereafter MR97). For an external shock,
the deceleration radius is $r_{dec} \sim (10^{16}
 {\rm cm}) \, (E_{51}/n_o)^{1/3} \Gamma_3^{-2/3}$, 
and the observed deceleration timescale 
$t_{dec} \sim r_{dec} /(c \, \Gamma^2)$, where $E_{51}$ is
the initial fireball kinetic energy, and
$n_o$ the particle  number density in the surrounding medium (we use 
the convention 
 $A_x = A/(10^x)$). We
 expect a temporal dependence of the type 
$\Gamma (t) \sim r^{-3/2} \propto t^{-3/8}$ and
$r_{\parallel} = c \, t \, \Gamma^2  \propto t^{1/4}$, 
with $r_{\parallel}$ the longitudinal radius.
If $I'_{\nu'}(t) $ is the (time-dependent) specific intensity
in the shock comoving frame ($\nu' = \nu/\Gamma$), the 
observer frame flux at a frequency $\nu$ is
then $F_{\nu} \propto r_{\perp}^2 \, \Gamma (t)^3 \, 
I'_{\nu}(t) \sim t^2 \, \Gamma (t)^5 \, I'_{\nu}(t)$ (MR97).
The temporal evolution  of $I'_{\nu}(t) $ depends on the adopted model of 
emission. 

The prompt burst high-energy behavior as observed in the 
1~keV -- 1~GeV  is  consistent with \syn
emission of impulsively accelerated particles  (Tavani 1996a,b,
hereafter T96a,b), 
possibly upscattered by inverse
Compton (IC) effects  (e.g., MR97).
We adopt here
the \syn model of  prompt GRB emission from a fireball,  and follow the evolution of 
the radiative shock as time progresses. IC  is not expected to play
an important role at the large values of $r_{\parallel}$
implied by 
the X-ray afterglow of \grbfp. A shock model of emission requires special
conditions in order to have detectable X-rays  8-80~hrs after a GRB.
Failed supernova (Woosley 1993), accretion-powered emission from massive
black holes (e.g., Woosley 1996) and spindown-powered emission from newly
 born spinars (e.g., Usov 1992) may also produce delayed X-ray emission.


Let us assume that most of the emission of an expanding and 
progressively decelerating fireball is produced by the reverse shock, 
i.e.,  by the ejecta particles 
cooled by adiabatic expansion. The comoving intensity 
at $r_{\parallel} > r_{dec}$ turns out to be $I'_{\nu'} \propto
n_{ej}' \, B' \, \Delta  r_{\parallel}'$, where $n_{ej}'$ 
is the ejecta radiating particle density, $B'$ the
magnetic field, and $ \Delta r_{\parallel}' = 
r_{\parallel}/\Gamma$. For adiabatic expansion, $n_{ej}'
\sim V'^{-1} \sim t^{-9/8}$,
$B' \sim V'^{-2/3} \sim t^{-3/4}$, and 
$\Delta r_{\parallel}' \sim t^{5/8}$, where we 
assumed a frozen-in magnetic field with
 $V' = 4 \, \pi \, r_{\parallel}'^2 \, \Delta r_{\parallel}'$
the comoving volume (MR97).
We therefore obtain $I'_{\nu'} \sim t^{-5/4}$, and 
$F_{\nu} \sim t^{-9/8}$.
The observed 
 X-ray afterglow can be 
residual \syn emission of non-thermal particles accelerated by the
prompt shock and adiabatically cooled. A non-thermal particle 
distribution function is assumed to be formed at the prompt GRB shock,
as indicated by broad-band GRB spectra (T96a,b).
This function is described by a relativistic
Maxwellian low-energy component below the 
critical energy $\gamma_*$, and by a power-law
component up to a maximum energy $\gamma_{max}$ above 
$\gamma_*$. The detectability of an X-ray afterglow from an expanding
fireball of this kind crucially depends on the \syn critical
frequency $\nu_* \propto \Gamma \, \gamma_*^2 \, B'$ evolved at the
radiating site. Since $\gamma_* \sim 1/r \sim t^{-1/4}$,
the critical frequency has a relatively fast decay, 
$\nu_*(t) \sim t^{-13/8}$. The observed X-ray flux
$f_x(t)$  is the integral of $F_{\nu}(t)$ 
over a fixed energy band (say, 2-10~keV), and it turns out to be 
$f_x(t) =  \xi_x (t) \,  \nu_* (t)  \, F_{\nu_*}(t)$,
 where $\xi_x (t)$ gives the fraction of total luminosity in the 
required energy range.
We obtain $f_x(t)  = \xi_x (t) \,  t^{-21/8}$, i.e., 
a decay behavior substantially steeper  than observed for \grbf
(but consistent with the lack of afterglow from \grbjp).

Shock emission from an adiabatically cooled fireball is predicted to
be non-thermal. However, due to the strong temporal  dependence  of $\nu_*$,
delayed
X-ray emission from impulsively accelerated particles is possible only
for very  efficient energization.
We can constrain the post-shock particle energization 
for no re-acceleration during adiabatic
expansion at $r_{\parallel} > r_{dec}$.
The observed  photon energy of the afterglow is assumed to be
$E_{max} \sim \Gamma (t)  \, \gamma^2_{max} (t) \, B' (t) \sim 
(t/t_*)^{-13/8}$.
The decay timescale $t_*$ can be deduced
from the time evolution of the peak energy of the $\nu \, F_{\nu}$ spectrum
during a typical GRB, 
$\nu_* \sim ({\rm 200 \, keV}) \, (t/t_*)^{-13/8}$.
A typical value is  $t_* \simeq 3$~s.
For no re-acceleration,
we obtain
\be  \gamma_{max}(0)/\gamma_*(0) \goe  0.22 \, (t/t_*)^{13/16} \label{eq-1} \en
 i.e., $ \gamma_{max}(0)/\gamma_*(0) \simeq 400$
for  $t = 8$~hrs. A substantial fraction of the shock luminosity
is required to be emitted in X-rays   even for values of
$\nu_*$ substantially smaller than $h \, \nu_* \sim 1$~keV.
If we adopt $F_{\nu} \sim \nu^{-1}$ above $\nu_*$
 (marginally consistent
with the OSSE observation of \grbf at $t = 0.5$~hrs, Matz \etal\ 1997),
we expect the fraction $\xi_x$ to be constant during  fireball evolution
as long as Eq.~1 is satisfied.
A steeper $F_{\nu}$ above $\nu_*$ would result in an  X-ray flux 
steeper than $f_x(t)  \propto t^{-21/8}$.
In this model, \syn emission peaks in the
optical range at $t/t_* \sim 1.8\times 10^3$. However, due to the
steep decrease of the bolometric luminosity, the delayed unabsorbed optical
emission from pure \syn is too faint to be detectable. 


An alternative model of fireball expansion relies on the emission by 
newly shocked material of  the surrounding  medium.
In this case, the density of the radiating particles of the comoving
frame is $n' = \Gamma \, n_o$.
In the absence of Rayleigh-Taylor instabilities at
the contact discontinuity, the forward shock fluid does not mix
with that of the reverse shock.
The magnetic field can be
deduced from equipartition arguments, $B' \simeq \lambda_B^{1/2} \, (8 \, \pi \,
n_o \, \Gamma \, \gamma_* \, m_p \, c^2)^{1/2}$, with 
$\lambda_B$ the efficiency  of
turbulent generation of magnetic field energy, and $m_p$ the proton's mass.
It is interesting that in this case,
$I'_{\nu'} \sim t^{-5/8}, F_{\nu} \sim t^0$, and
$\nu_* \sim t^{-3/2}$ (e.g., MR97).
We therefore deduce the decay behavior of the bolometric \syn emission in
the 2-10~keV band as $f_x(t) = \xi_x (t) \, t^{-3/2}$.
For a relatively constant $\xi_x (t)$, this function is
marginally consistent
with the observed afterglow of \grbfp.
As for  the previous model, we can obtain the conditions on the
acceleration mechanism that make possible delayed X-ray emission.
If the local $\gamma_* \sim \Gamma$, and $\gamma_{max} \sim \alpha' \,
\Gamma$, with $\alpha' \goe 1$, we can translate the condition on
the observed photon energy as a requirement on initial values of 
$\gamma_*$ and $\gamma_{max}$. For $E_c = 10$~keV and $t=$~8~hrs, we obtain
$\Gamma_o^4 \sim 4.7 \times 10^{18}$, i.e, $\Gamma_o \simeq 4.6 \times 10^4$.
More  generally, we obtain for no re-acceleration
\be \gamma_{max} (0) ^2 \, \gamma_* (0)^{1/2} \simeq 4.7 \times 10^{15} \,
n_o^{-1/2} \, \Gamma_{o,2}^{-3/2} \, \lambda_B^{-1/2} \,
(t/{\rm 8~hrs})^{3/2}      \label{eq-2} \en
We deduce that a forward shock may be consistent with the
observed afterglow of \grbfp, but only if the particle acceleration process
of the forward shock material is very efficient and/or
$\Gamma_o$ is very large.
Fig.~1 show the results of the calculated  flux in the X-ray and
optical R bands  for two different choices of the power-law index of
the post-shock particle distribution function corresponding 
to $F_{\nu} \sim \nu^{-1}$
and $F_{\nu} \sim \nu^{-3/2}$, respectively.
From Fig.~1 we deduce that in the absence of additional acceleration,
the predicted X-ray flux does not agree  with observations of
\grbf if $t_* \loe 100$~s.
The resulting  optical transient  emission in the absence of absorption
effects  is shown in Fig.~1 (dot-dashed curves).
The initial ratio of optical to X-ray flux is $\loe 10^{-5}$,
implying an optical transient of R-magnitude $m_R \sim 18.4$.
The transient evolves as $t^{-3/2}$ up to $t/t_* \sim 3\times 10^3$
corresponding to the time when $\nu_*$ sweeps the optical band.
For later times, the evolution of the optical flux becomes  steeper.
Detectable delayed radio emission may also be expected in this class of models
(Paczy\'nski \& Rhoads 1993).

If particle re-acceleration occurs during the adiabatic expansion phase,
the constraints on $\gamma_{max} (t) $ and $ \gamma_* (t)$
become less stringent.
Turbulent mixing of the reverse and forward shock fronts may favor
efficient particle energization. In this case, Eq.~\ref{eq-2} may not
apply and a milder constraint for $\gamma_{max} (t) $ and $ \gamma_* (t)$
can be required. The evolution of both the X-ray and optical fluxes can be less
steep than calculated here.


Models based on relativistic jets from failed supernovae (Woosley 1993)
are also based on forward shock emission. Depending on the density of the
surrouding medium, dissipation by \syn (or bremsstrahlung) emission
can result in different  values of the decay constant $t_*$ compared
to models considered above. However,  the qualitative features of the X-ray
evolution should follow the trend given above.
We can deduce an interesting constraint on the `spinar' model
for GRBs.
A newly-born strongly magnetized neutron star can spin down very
rapidly and make possible  an initial  fireball and
subsequent non-thermal `pulsar-like' emission (e.g., Usov 1992).
In order to explain the required energy near $10^{52} \ergs$, this
model assumes extreme values of the surface magnetic field
$\sim 10^{15}$~G and initial spin period  $P \loe 10^{-3}$~s.
The spindown energy varies with time as $\sim (t/t_{sd})^{-2}$,
with $t_{sd}$ the spindown timescale, $t_{sd} \sim \dot{P}/P\sim 100$~s.
A fraction of the total spindown luminosity can be emitted as high-energy
radiation.
At $t = 8$~hrs, the emitted luminosity is 
decreased by a factor of $\sim 10^{-5}$ compared to the initial one.
At $t = 85$~hrs, the decrease factor is
$\sim 10^{-7}$. Therefore,
this model predicts a luminosity evolution 
in disagreement with the observations of \grbfp.

\section{Galactic models}
 
At a distance $d = d_{100} \, 100$~kpc, an observed X-ray
flux $F_x = F_{-12} \, 10^{-12} \flux$ corresponds to an
isotropically radiated luminosity,
$L_x \simeq 10^{36} \, F_{-12} \, d_{100}^2 \rm erg \, s^{-1}$.
This sub-Eddington luminosity for a solar mass compact star
can be produced in different ways.
%
%
A sudden internal explosive event from a neutron star might
cause matter liftoff and substantial release of energy in the
external crust.  Neutron stars in an extended Galactic halo
require explosion energies of order of $10^{41}-10^{42}$~ergs.
A super-Eddington luminosity leads to matter liftoff from the
surface, possibly followed by  matter fallback  onto the
surface of the neutron star.
Let us consider first the case of internal energy release
and cooling with no contribution from subsequent accretion.
The thermal response of a neutron star crust  depends on
the depth at which most of the residual burst energy is liberated.
For deep energy deposition,  the timescale for conduction-driven
cooling might be of orders of years or more.
On the other hand, energy deposition in shallow layers of the external crust
may lead to cooling timescales of order of $10-10^4$~s.
This can be seen  by considering the initial photon diffusion
time of the surface layers of a neutron star, $\tau_{diff} \sim
3 \, \kappa \, \rho \, x^2 \, u_{th} / (c \, u_{\gamma})$,
where $\kappa$ is the radiative opacity, $\rho$ the mass density,
$x$ the depth (in cm), $u_{th}$ the thermal energy, $c$ the speed of light,
and $u_{\gamma}$ the radiation energy density (e.g., Eichler \& Cheng 1989).
For $^{56} \rm Fe$ matter, we obtain
$\tau_{diff} \simeq ({\rm 30 \, s}) \, (x/100)^2 \, (T_8 / 0.3)^{-5.5} \,
\rho_5^{2.4}$. 
Thus  a post-burst neutron star surface radiating near 
the Eddington limit can have $T_8 \sim 0.3$ and $\tau_{diff} \sim 30$~s.
Solutions of the conductive heat flow equation for
vanishing depths  depend on how
the heat conductivity and heat capacity vary as a function of depth.
This dependence is poorly known and we consider here the general form
of the solution $T(t) = T_o \, (t/\tau_c)^{-1 + \zeta}$,
where $\zeta$ depends on the functional dependence  of heat conductivity and
capacity on depth.

We can then assume that the post-burst  bolometric flux
from a neutron star surface  as detected by a distant observer depends on time as
$L(t') = L_o \, t'^{-\alpha}$. This  implies  an effective  surface temperature
dependence $T(t') = T_o \, t'^{-\alpha/4}$,
where $t' = t/\tau_c$ and $T_o \propto L_o^{1/4} \,
(1 - R_{\rm Sch}/R_*)^{-1/4}$,
with $R_*$ and $R_{\rm Sch}$ the neutron star and Schwarzschild radii,
respectively.
For an initial luminosity radiated by the whole star  near  the
Eddington limit $L_E = L_{38} \, 10^{38} \ergs$,
we deduce an initial temperature, $T_o \simeq 2\times 10^7
\, L_{38}^{1/4}$~K. It is interesting that for a decay timescale 
 $\tau_c \sim  100$~s and exponent
$\alpha = 3/2$, 
the temperature of a cooling
neutron star surface would  be 
 in an optimal range  for detection by X-ray instruments
$\sim$1 day after the burst.
Fig.~2 shows the calculated energy flux in the 2-10 keV
energy band  in units of the initial  bolometric  flux emitted by a
cooling  neutron star decaying as $f_x(t') = f_o \,
t'^{-3/2}$, where $f_o = L_o/(4 \, \pi \, d^2)$. 
We consider two models characterized by different initial temperatures,
$T_o = 2\times  10^7$~K and $T_o = 3\times  10^7$~K, respectively.
We notice that the X-ray fluxes detected 
approximately 8 and 85 hours following \grbf qualitatively  agree with
expectations based on this simple model.
An  interpretation
in terms of thermal afterglow implies $\zeta \simeq 5/8$,
a value that can be used for a detailed modelling of
 the  thermal response of a neutron star.

\begin{table*}
\begin{center}
TABLE 1: Models for X-ray afterglows from GRBs
\vskip .04in
\begin{tabular}{lll}
\hline
Model                      & $f_x(t)  $  &   Requirements\\
\hline
Adiabatic fireball: reverse shock & $ \xi_x \, t^{-21/8}$  & $\xi_x \sim$~const.;
$ \gamma_{max}(0)/\gamma_*(0) \goe 400$\\
Adiabatic fireball: forward shock & $\xi_x \, t^{-3/2}$  & $\xi_x \sim$~const.;
$\gamma_{max} (0) ^2 \, \gamma_* (0)^{1/2} \goe 4.7 \times 10^{15} \,
n_o^{-1/2} \, \Gamma_{o,2}^{-3/2} $\\
Failed supernova                          & $\xi_x \, t^{-3/2}$ & 
Optically thin circumstellar medium\\
Spindown-powered emission                 & $t^{-2}$  & Strongly magnetized, rapidly rotating compact star\\
NS cooling 
 & $t^{-1 + \zeta}$  & Shallow energy deposition,
$\zeta \sim 5/8$\\
NS cooling  with external irradiation 
   &  $t^{-1 + \zeta'}$  & Substantial reprocessing, $\zeta' \sim 5/8$\\
\hline
\end{tabular}
\end{center}
\end{table*}

The cooling neutron star model predicts a substantial
softening  of the spectrum.
 Fig.~2  also shows  the  calculated evolution of the 
hardness $H$ defined as the ratio of photon fluxes
in the 3-6~keV and 2-3~keV bands.  
The hardness ratio is predicted to  vary by a factor $\sim 2$
within the first SAX TOO observation of \grbfp.
An even more drastic variation of the hardness dropping below $0.1$ is
predicted for later times, e.g., during the second SAX TOO observation.
%
No detectable optical emission is expected  in this case from
pure neutron star cooling.


Other ways of producing X-ray emission can be considered.
If the bulk of GRB emission occurs outside but not too far  from a
neutron star, photon irradiation and particle precipitation
can further heat  its surface.
The energy deposition from above can lead to a thermal relaxation
qualitatively similar to that of a cooling neutron star surface heated
from below. However, the detailed   heat transport may be  different than
in the previous case.
%
%
Furthermore,
fallback material from the burst explosion can settle in a
disk or spherical inflow onto the neutron star surface.
The X-ray  spectrum in this case is predicted to be a combination of
blackbody and power-law components as observed from
several low-luminosity accreting neutron stars
(White, Nagase \& Parmar 1995).
X-ray pulsations might  be detectable in case of emission from a strongly
magnetized neutron star.
The ratio of optical to X-ray luminosity is expected to be
$10^{-2}-10^{-3}$, i.e., similar to  faint
accreting compact sources (e.g., White et al. 1995).


\section{Discussion and conclusions}

We showed  
that the detection of an X-ray
afterglow  from  a  GRB strongly   constrains   theoretical models.
Table~1 summarizes the main properties of the models considered here.
Cosmological models based on adiabatic evolution of a \syn radiating
shock front are the most constrained.
The calculated  emission 
  in models of adiabatically expanding  reverse shocks
does   not agree with  observations of \grbfp.
Forward shock models  might be compatible with observations 
only for very efficient particle acceleration, large decay timescale
$t_* \goe 100$~s or  re-acceleration during expansion.
An optical transient of initial  magnitude  $m_R \sim 18-19$  lasting
a few hours after  the GRB is expected in this model.
If $t_* = 300$~s, we deduce from Fig.~1 an X-ray flux in qualitative agreement with
the observations of \grbf and an optical magnitude at $t/t_* \sim 100$ equal to 
$m_R \sim 22.7$, i.e., slightly higher than the limiting magnitude ($m_R \simeq 22$)
of the observations at $t=15.3$~hr reported by Guarnieri \etal\ (1997).
Consistency  with the optical transient possibly associated with \grbf
(Groot \etal\ 1997) requires $t_* \goe 300$~s.
 No radio transient source above $\sim $1~mJy 
has been reported in searches at 5~GHz of  the refined   error boxes of
 \grbj and \grbf (Frail \etal\ 1997, Galama \etal\ 1997).
%
%
%
We also showed that
 Galactic models of  post-burst emission from cooling neutron stars
have definite predictions. 
Different flux and spectral evolution patterns  of
X-ray afterglows are
expected as a function of initial surface temperature and cooling timescale.

Our analysis is important for the interpretation of 
follow-up  X-ray observations of \grbfp.
As  shown in Fig.~2, in neutron star cooling models 
no detectable X-ray flux above $10^{-14} \flux$ from
\grbf  is expected in the X-ray band for $t \goe 100$~hrs.
Analogously, cosmological  models  based on forward shocks with no re-acceleration,
predict an evolution of the  X-ray flux
  $f_x \sim t^{-3/2}$ or steeper. Again, the predicted flux at
$t \goe 100$~hrs is below detectability with current instruments.
A violation of these predictions would be remarkable, since it
would point to a steady-state high-energy emission of the GRB
  counterpart.

We emphasize that  X-ray afterglows from GRBs show different time evolutions.
 Within a given model, variations of  physical
properties (critical timescales, temperature, maximum value of particle 
energies in non-thermal models, etc.) can explain the difference
between the afterglows from  \grbj and \grbfp.
A combined sequence of fast  multiwavelength observations following
GRBs can further constrain the emission parameters and their variations
among different GRB sources.

\vskip .1in
The author thanks E. Costa, F. Frontera,  M. Ruderman and
M. Feroci for   discussions.
Research supported in parts by the NASA grant NAG5-2729.

\newpage

\baselineskip 15pt
\centerline{\bf References}

\rref{Band, D., \etal, 1993, ApJ, 413, 281}


\rref{Boller, T. \etal, 1997, IAU Circular no. 6580}

\rref{Butler, R.C., \etal, 1997, IAU Circular no. 6539}

\rref{Costa, E., \etal, 1997a, IAU Circular no. 6533}
\rref{Costa, E., \etal, 1997b, IAU Circular no. 6572}
\rref{Costa, E., \etal, 1997c, IAU Circular no. 6574}

\rref{Eichler, D. \& Cheng, A.F., 1989, ApJ, 336, 360}

\rref{Frail, D.A. \etal, 1997, IAU Circular no. 6576}

\rref{Galama, R., \etal, 1997, IAU Circular no. 6574}

\rref{Groot, P.J., \etal, 1997, IAU Circular no. 6584}

\rref{Guarnieri, A., \etal, 1997, IAU Circular no. 6582}

\rref{Hurley, K. \etal, 1997a, IAU Circular no. 6571}

\rref{Hurley, K. \etal, 1997b, IAU Circular no. 6578}

\rref{in 't Zand, J., \etal, 1997, IAU Circular no. 6569}

\rref{Matz, S. \etal, 1997, IAU Circular no. 6578}


\rref{M\'esz\'aros, P. \& Rees, M.J., 1997, ApJ, 476, 232}

\rref{Paczy\'nski, B. \& Rhoads, J.E., 1993, ApJ, 418, L5}

\rref{Rees, M.J., 1966, Nature, 211, 468}

\rref{Tavani, M., 1996, PRL, 76, 3478}
\rref{Tavani, M., 1996b, ApJ, 466, 768}

\rref{Usov, V., 1992, Nature, 357, 472}
 
\rref{White, N.E., Nagase, F. \& Parmar, A.N. 1995, in
{\it X-Ray Binaries}, eds. W.H.G. Lewin, J. van Paradijs \& E.P.J.
van den Heuvel (Cambridge University Press), p. 1}

\rref{Woosley, S. E., 1993, ApJ, 405, 273}

\rref{Woosley, S. E., 1996, in 
{\it Gamma-Ray Bursts}, eds. C. Kouveliotou, M.F. Briggs, G.J. Fishman
(AIP Conf. Proc. no. 384), p. 709}

\baselineskip 14pt


\newpage
\vspace*{-.9in}
\epsfxsize=6.5in
\epsfysize=7.5in
\epsffile{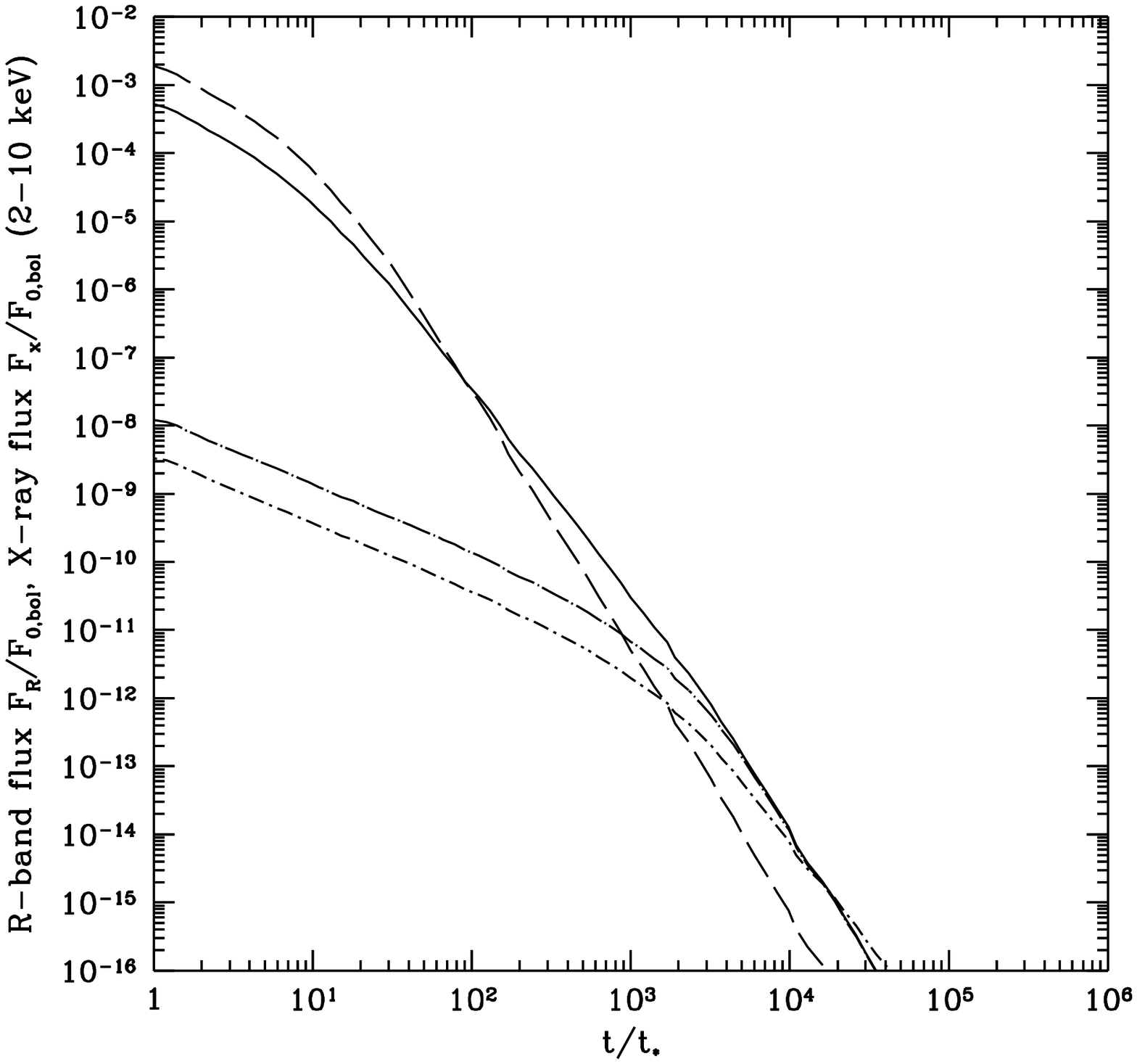}
\vspace*{-.1in}
\nn
{\small FIG.~1 --
Calculated evolution of
 X-ray and optical (R-band) fluxes  for an adiabatically expanding
forward shock with no re-acceleration.
The temporal axis scale is in units of $t_*$, with  the \syn
critical frequency evolving  as $\nu_* \sim (t/t_*)^{-3/2}$.
{\it (Solid curve:)}  X-ray
(2-10 keV) flux (in units of the initial bolometric flux)
 for a spectral form $F_{\nu} \sim \nu^{-1}$ above $\nu_*$;
{\it (long-dashed curve:)}  the same  for
$F_{\nu} \sim \nu^{-3/2}$ above $\nu_*$.
{\it (Long-dashed-dotted curve:)} 
R-band flux (in units of the initial bolometric flux)
for  $F_{\nu} \sim \nu^{-1}$ above $\nu_*$;
{\it (short-dashed-dotted curve:)}
the same  for
$F_{\nu} \sim \nu^{-3/2}$ above $\nu_*$. }
 
\newpage
\vspace*{-.9in}
\epsfxsize=6.5in
\epsfysize=7.5in
\epsffile{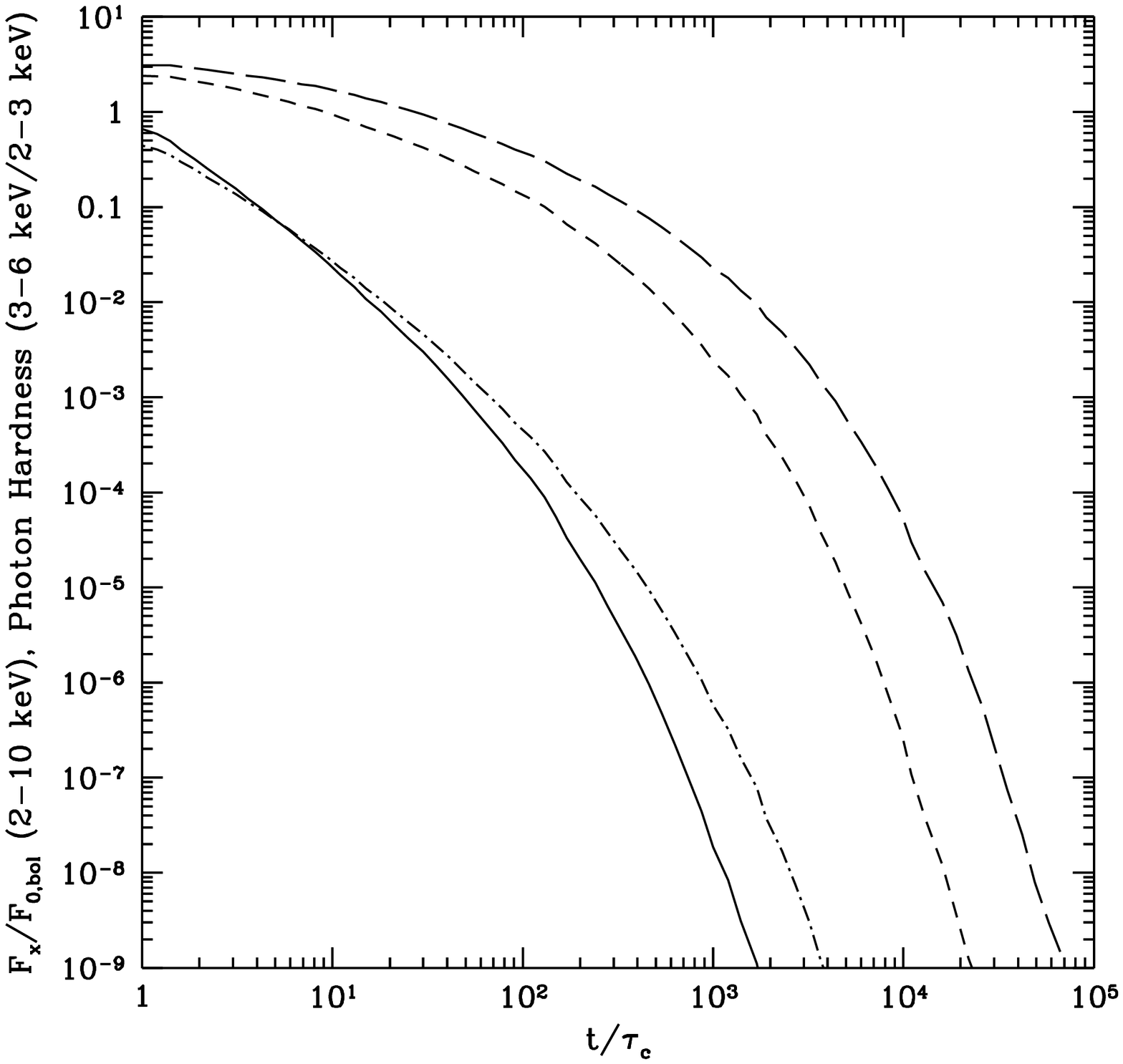}
\nn
{\small FIG.~2 -- 
Evolution of X-ray flux and photon hardness
as a function of $t/\tau_c$ for the cooling neutron star model.
{\it (Solid curve:)}   X-ray flux
(2-10 keV) in units of $F_{o,bol}$ for an initial temperature $T_o = 2\times
10^7$~K;
{\it (dashed-dotted curve:)}
the same for $T_o = 3\times  10^7$~K.
{\it (Short-dashed curve:)} photon hardness ratio for the channels
3-6 and 2-3~keV for $T_o = 2 \times  10^7$~K;
{\it (long-dashed curve:)} photon hardness ratio for $T_o =  3\times 10^7$~K.}

\end{document}